\documentclass[aps,prl,showpacs,twocolumn,groupedaddress]{revtex4}
\usepackage[latin1]{inputenc}
\usepackage{graphicx}
\usepackage{color}

\begin{document} 


\title{Behavior of Optical Phonons near the Diffuse Phase Transition 
in Relaxor Ferroelectric PbMg$_{1/3}$Ta$_{2/3}$O$_3$}



\author{S.G. Lushnikov}
\email[]{sergey.lushnikov@mail.ioffe.ru}
\affiliation{Ioffe Physical Technical Institute, 26 Politekhnicheskaya,
194021, St. Petersburg, Russia}
 
\author{S.N. Gvasaliya}
\altaffiliation{On leave from Ioffe Physical Technical Institute, 
26 Politekhnicheskaya, 194021, St. Petersburg, Russia}
\affiliation{Laboratory for Neutron Scattering ETHZ \& Paul-Scherrer
Institut CH-5232 Villigen PSI Switzerland} 

\author{R.S. Katiyar}
\affiliation{University of Puerto Rico, P.O. Box 23343, San Juan, USA}

\date{\today}

\begin{abstract} 
Raman scattering in relaxor ferroelectric PbMg$_{1/3}$Ta$_{2/3}$O$_3$ (PMT) 
was investigated in the single crystalline form in the temperature range of 
20~-~295~K. Anomalous temperature dependence of the integrated intensity and 
the Raman line contours were found at the diffuse phase transition. A 
correlation between the anomalies in the integrated intensities and the 
dispersion of the dielectric response was observed. The distortions of Raman 
lines with decreasing temperature are discussed.
\end{abstract}

\pacs{77.80.-e, 61.12.-q} 

\maketitle

\noindent  

The ferroelectric state in partially disordered crystals has attracted considerable 
attention of researchers due to their anomalously high dielectric constant over a 
broad temperature range near the phase transitions. The most suitable model objects 
for the studies of the ferroelectric state in partially disordered crystals are complex 
perovskites with the common formula AB$'_x$B$''_{1-x}$O$_3$~\cite{smol1}. In some 
compounds of this family, the transition into a ferroelectric phase drastically differs 
from the classical ferroelectric phase transitions observed in perovskites and manifests 
itself as wide (several hundred degrees) frequency-dependent anomalies of physical 
properties referred to as a diffuse phase transition~\cite{smol1}. This group of compounds 
is called relaxor ferroelectrics, or simply relaxors~\cite{cross}. At first it was 
believed that such an anomalous lattice dynamics is due to composition fluctuations 
accompanied with or without local phase transitions~\cite{smol1}. Later it was found 
that the relaxor ferroelectrics are homogeneous, but have nanoregions with 1:1 
ordering of B$'$ and B$''$ ions~\cite{burns} that play an important role in lattice 
dynamics of complex perovskites. 

In relaxors, the transformation into a ferroelectric phase can be accompanied 
by a first-order structural phase transition. In some cases the structural instability 
of a crystal is pronounced in lattice dynamics, as, for instance, in relaxor ferroelectric 
PbSc$_{1/2}$Ta$_{1/2}$O$_3$~\cite{smol1}. In other cases, it can be suppressed, and 
then it is realized only under an applied external electric field as, for instance, 
in PbMg$_{1/3}$Nb$_{2/3}$O$_3$ (PMN)~\cite{smol1}. Probably, this feature does not 
exert a considerable influence on lattice dynamics in the paraelectric phase, 
but it should be taken into consideration in studies of the nature of the ferroelectric 
relaxor state. The most suitable object for studies of the transformation into 
the relaxor state is PbMg$_{1/3}$Ta$_{2/3}$O$_3$ (PMT) - an analog of PMN~\cite{smol1}. 
The lattice dynamics of the PMT crystal has been investigated by dielectric~\cite{ko} 
and optical~\cite{pis,siny1} spectroscopy, neutron scattering~\cite{seva1,seva2}, and 
adiabatic calorimetry~\cite{moriya} in a wide temperature range, but, to our knowledge, 
no anomalies except the relaxor one have been revealed even under an applied electric field. 
The transformation into the ferroelectric phase in PMT is accompanied by a broad 
frequency-dependent anomaly of the dielectric response with a maximum at a frequency 
of 10 kHz in the vicinity of 170 K, which is well described by the Vogel-Fulcher law~\cite{ko}. 
In the entire range of temperatures and applied electric fields, the macroscopic symmetry of 
PMT does not vary and remains cubic Pm$\bar{3}$m~\cite{smol1,lu}.

In experimental studies of lattice dynamics of PMT, polarized first-order Raman scattering 
was observed~\cite{pis,siny1}, which contradicts X-ray structural investigations~\cite{smol1,lu} 
because first-order Raman scattering is forbidden by the selection rules for cubic crystals 
with the perovskite structure. It is important to note that first-order Raman scattering 
was observed for nearly all complex perovskites~\cite{siny2}.  The nature of this scattering 
is a subject of discussion; the only commonly accepted point of view is that Raman scattering 
and the short-range order in distribution of B$'$ and B$''$ ions (in our case, Mg and Ta) 
are related. Different ideas on the nature of Raman scattering in complex perovskites were 
analyzed in detail in~\cite{siny2,siny3}, and we shall not dwell on them here. Note only 
that Raman spectra of a major part of complex perovskites are similar to each other. 
      
Raman scattering studies of lattice dynamics of complex perovskites were performed in detail 
in the paralectric phase~\cite{siny3}. The evolution of low-frequency Raman spectra 
of PMN, a model object for studies of relaxor ferroelectrics, was investigated in an 
unsuccessful search for soft modes in a wide temperature range~\cite{siny2,siny3,siny4,toulouse}. 
There is considerably less information on the behavior of optical phonons in the vicinity of 
the diffuse phase transition. In the work reported here we carried out Raman scattering 
investigations of the PMT crystal to obtain data on the behavior of optical phonons in 
the case of a purely 'relaxor behavior' of the lattice dynamics.

Raman spectra were excited with an argon laser and analyzed with a triple grating spectrometer 
ISA Model T64000 equipped with a liquid nitrogen cooled CCD detector. The backscattering 
spectra were recorded using a Raman microprobe system. The illumination of the sample was 
adjusted before measurements at each temperature to optimize the signal. Measurements were 
performed using a modified Cryogenic Tech. Closed-cycle Refrigerator Model 20 with a Lake 
Shore DRS-84C temperature controller. The diagonal X(YY)$\rm\bar{X}$ and off-diagonal 
X(YZ)$\rm\bar{X}$ spectra were collected with X, Y and Z- axes being along the four fold 
directions of the PMT cubic lattice. When discussing spectra and assigning the lines, we 
adopted the notations used in~\cite{siny1}. Raman measurements were carried our on a 
high-quality single-crystal of PMT $6.5\times4.5\times1.6$ mm$^3$ in size. The same single 
crystal was used for neutron scattering~\cite{seva1} and Brillouin light scattering~\cite{ko} 
experiments. No changes in the crystal quality were observed upon thermal cycling. 
%
\begin{figure}[h]
  \includegraphics[width=0.5\textwidth, angle=0]{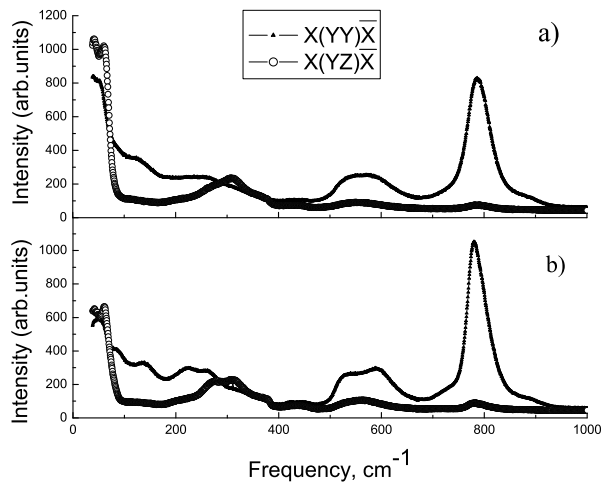}
  \caption{The diagonal and off-diagonal Raman spectra of PMT T=285 K (a) and T=134 K (b). 
Note the shape deformation of high frequency A1g mode at lower temperature.}
\label{fig1}
\end{figure}

Figure 1 shows Raman spectra of the PMT crystal for both diagonal and off-diagonal polarizations 
at two temperatures. The off-diagonal X(YZ)$\rm\bar{X}$ spectra consist of two modes with 
characteristic frequencies 80~cm$^{-1}$ and 300~cm$^{-1}$. Due to similarity with the spectrum 
of complex perovskites~\cite{siny1,siny2}, it was concluded that these modes are of the T$_{2g}$ 
symmetry and are related to the Pb and O motion. The diagonal X(YY)$\rm\bar{X}$ spectra are richer - 
they consist of four modes with 80, 240, 580, and 800~cm$^{-1}$. As pointed out 
in~\cite{siny1,siny2}, the intense line in the diagonal spectra at 800~cm$^{-1}$ is the 
A$_{1g}$ mode. It represents the breathing-type motion of oxygen ions in the octahedral. 
There are two significant changes in the PMT spectra as the temperature decreases below 
room temperature: a strong decrease of the intensity below 200~cm$^{-1}$ 
and the A$_{1g}$ mode asymmetry. 

Experimental results were analyzed by using calculations of integrated reduced Raman scattering 
intensity given by:  
\begin{equation}    
\label{intensity} 
S=\int_{\omega_1}^{\omega_2}\frac{I}{n(\omega)+1}d\omega,
\end{equation}   
where $\omega_1$ and $\omega_2$ are the boundaries of Raman bands and $n(\omega)$ is the Bose 
population factor. The integrated intensity is proportional to susceptibility of a 
relevant mode. This approach is more correct for analysis of the temperature behavior of Raman 
spectra of the crystals whose line shapes are considerably distorted and damping is high. 
Such an analysis was successfully used for treatment of Raman measurements in 
a disordered perovskite KTa$_{1-x}$Nb$_x$O$_3$~\cite{fontana}. It is not difficult to 
calculate the integrated intensity for the high-frequency A$_{1g}$ mode in the PMT crystal, 
because it is well isolated in frequency from other vibrations and the integration interval is 
easily determined (Fig. 1). The situation with finding integration intervals for 
the low-frequency T$_{2g}$ mode is much more complicated, in particular because of its proximity 
to the exciting line. Therefore, the choice of integration limits for calculations of 
the integrated intensity of the T$_{2g}$ mode is, to a certain extent, arbitrary. The 
calculations performed for different integration limits have shown that qualitatively the 
results are the same. Only quantitative changes in the obtained anomaly (of the order of 3$\%$) 
can occur. Figure 2 shows the temperature dependence of integrated reduced Raman scattering 
intensity of A$_{1g}$ and T$_{2g}$ modes. It can be seen from Fig. 2 that the integrated 
intensity of both A$_{1g}$ and T$_{2g}$ lines below 80~K and above 220~K is nearly 
temperature-independent, which corresponds  to the predicted behavior of the first-order 
Raman spectrum of the crystal with temperature. The temperature dependence from 90 to 220~K is 
much more complicated: a decrease in temperature from room temperature at first leads 
to a reduction in intensity, which is pronounced for both modes, and then a sharp rise 
followed by a smooth fall to the temperature-independent portion of the dependence occurs. 
Note that the rise in the integrated intensity $\Delta$S for the high-frequency A$_{1g}$ mode 
is approximately 25$\%$ of its initial value (see Fig. 2a). This demonstrates once more 
how nontrivial the behavior of the vibrational spectrum of a relaxor ferroelectric is. It can 
hardly be expected that the high-frequency mode in the ferroelectric crystal can experience 
such large changes in the vicinity of the dielectric anomaly, especially in the absence 
of a structural phase transition. 
%
\begin{figure}[h]
  \includegraphics[width=0.5\textwidth, angle=0]{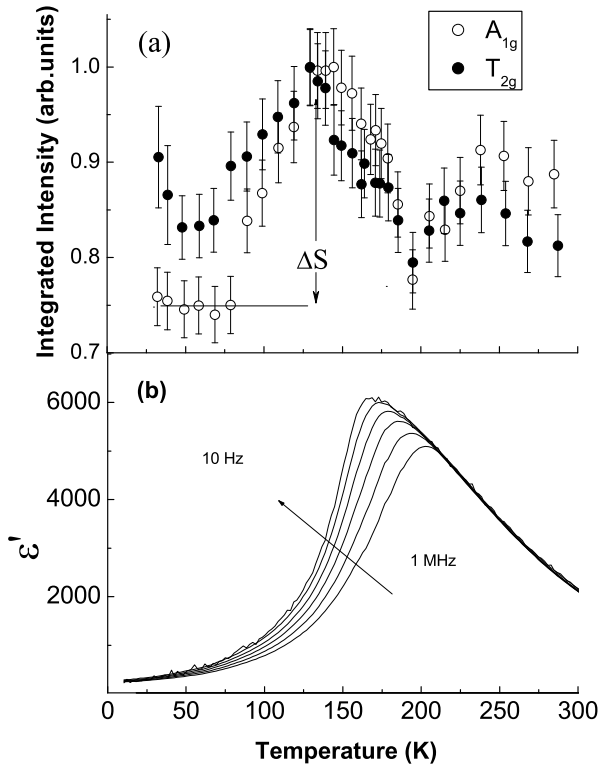}
  \caption{Fig.2a Temperature dependences of the reduced integrated Raman scattering 
    intensity of T$_{2g}$ and A$_{1g}$ modes. Fig.2b The real part of the dielectric 
    constant of PMT (data taken from~\cite{ko}).}
\label{fig2}
\end{figure}

It is interesting to compare the behavior of the Raman spectra of PMT and BaMg$_{1/3}$Ta$_{2/3}$O$_3$ 
(BMT), i.e., the compound with a low dielectric constant $\varepsilon'\sim20$~\cite{ko} 
where there is no phase transformation into the ferroelectric state. As noted above, the 
Raman spectra of PMT and BMT crystals are very similar (with the exception of the line widths), but 
their behavior with temperature differs considerably. As the temperature decreases 
(from the paraphase $\sim$ 1100 K), the first-order Raman scattering spectra of BMT and PMT arise 
in the vicinity of 900~K. With further decrease in the temperature, the behavior of the Raman 
spectrum of BMT depends upon the temperature only slightly (the population factor is taken 
into account). The temperature dependences of the integrated intensities of the A$_{1g}$ mode 
and other modes in BMT Raman spectra have no other anomalies except the high-temperature one~\cite{siny5}, 
while Raman spectra of  PMT crystals exhibit an anomalous behavior in the vicinity of the diffuse phase 
transition. 

In the case of normal ferroelectric perovskites much attention has been paid to the soft mode 
dynamics. In the absence of a soft mode, anomalies in the temperature dependences of the integrated 
intensity of hard modes in Raman spectra in the vicinity of the phase transitions can 
be expected~\cite{bis,ginzburg}. These will be due to different types of interaction between 
fluctuations of dielectric permeability and order parameter~\cite{ginzburg}. However, the formalism 
suggested in~\cite{ginzburg} is not suitable for discussing anomalies in the behavior 
of Raman spectra of PMT crystals where there is no structural phase transition. 
%
\begin{figure}[h]
  \includegraphics[width=0.5\textwidth, angle=0]{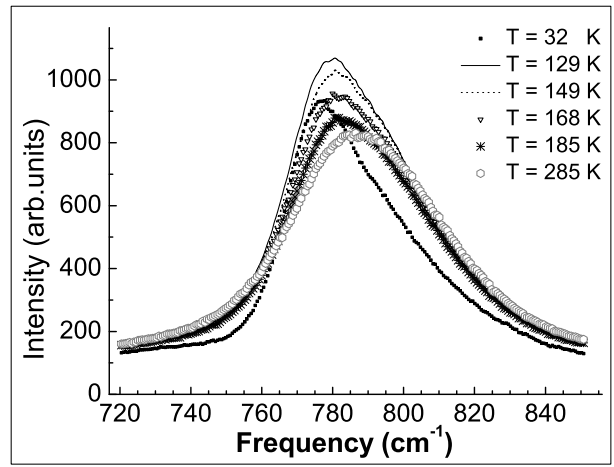}
  \caption{Temperature evolution of A$_{1g}$ mode of PMT crystal in the 
   vicinity of broad dielectric anomaly.}
\label{fig3}
\end{figure}

Let us consider in detail the anomalous behavior of Raman spectra of PMT in the vicinity of 
transformation into the ferroelectric state by comparing them with the behavior of the dielectric 
response (we compare Fig. 2a and 2b). It may be concluded that i) the susceptibility of hard modes 
of the PMT crystal does not depend upon the temperature when there is no dispersion of dielectric 
permeability (it is implied that the dispersion region is somewhat wider than that shown in Fig. 2b), 
ii) the anomaly in the integrated intensities of Raman modes is observed just in the region 
of dispersion. The commonly accepted mechanism responsible for the behavior of the real 
part of dielectric constant in relaxor ferroelectrics is a wide spectrum of relaxation times 
which is frozen in the PMT crystal at T$_f$ = 124~K. This temperature was determined in~\cite{ko} 
with the help of the Vogel-Fulcher law. In Fig. 2a, the maxima in the integrated intensities of 
both A$_{1g}$ and T$_{2g}$ modes are reached at $\rm T\sim124$~K. It is not likely that this 
coincidence in temperatures is accidental. Probably, the relaxation processes giving rise 
to the dispersion of dielectric response are responsible for the anomalous behavior of hard 
modes in the Raman spectra of PMT as well. While the existence of connection between 
relaxation processes and the low-frequency T$_{2g}$ mode is predictable, the connection 
between the high-frequency ($\sim790$~cm$^{-1}$) A$_{1g}$ mode and the relaxation processes 
is absolutely unexpected. Note that a considerable change in the intensity of 
diffuse neutron scattering in the PMT crystal was also observed in the temperature range 
of 80 - 220~K~\cite{seva1}.

Special attention should be paid to the shape of the high-frequency A$_{1g}$ mode that shows 
a nontrivial temperature evolution. It is evident from Fig. 3 that a decrease in temperature 
from room temperature results in distortion of the line shape. A wide but rather symmetric 
line shape in the paraelectric phase of the PMT crystal is distorted with decreasing temperature, 
becoming more symmetric and its high-frequency edge is extended on approaching 32~K. A 
nontrivial temperature evolution of the contour of the A$_{1g}$ mode in relaxors was 
observed for the first time for the PbSc$_{1/2}$Ta$_{1/2}$O$_3$ (PST) crystal in~\cite{siny5}. 
However, evolution of the A$_{1g}$ mode contour in the paralectric phase in PST was 
accompanied by its splitting and formation of an additional structure with increasing 
temperature. This behavior of the A$_{1g}$ mode was attributed to the transition into an 
intermediate (probably, incommensurate) phase in the PST crystals and dynamic 
breaking of the selection rules. Due to this, the A$_{1g}$ mode can be observed from other 
points of the Brillouin zone in light scattering experiments. Taking into account the 
X-ray and neutron diffraction data for PST, this interpretation is correct. For PMT crystals, 
the analysis of the X-ray and neutron diffraction data~\cite{lu} did not reveal the existence 
of additional phases. Note also that in PMT crystals the distortion of line contours is 
not accompanied by the formation of an additional structure. Therefore, it is reasonable to 
suppose that there is a connection between the distortion of the A$_{1g}$ line and an increase 
in the anharmonicity of the PMT crystal at low temperatures. This anharmonicity might 
be related to the formation of short-range order in PMT at low temperature. Indeed, 
the evolution of the neutron diffuse scattering in PMT crystal was observed at 
similar temperatures~\cite{seva1}. Attempts to describe Raman light scattering spectra 
by decomposing the spectra into the sum of several damped oscillators or Lorentz (Voigt) 
functions are not physically justified. In the absence of structural phase 
transitions, in order to describe the line contour it is necessary to introduce 
a larger number of peaks at low temperatures than at high ones, and their number is unlimited. 

In summary, a detailed study of Raman scattering spectra of relaxor ferroelectric PMT crystal in 
the vicinity of the ferroelectric transition temperature has revealed the following:

* There is correlation in the behavior of the optic hard mode and the dielectric response in 
the vicinity of the diffuse ferroelectric phase transition indicating that the relaxation processes 
make contributions to the temperature evolution of the hard modes. This correlation and the anomalous 
behavior of the hard modes were observed for the first time; 

* The distortion of the high-frequency A$_{1g}$ mode, in the absence of a structural phase transition, 
is attributable to an increase in the crystal anharmonicity with decreasing temperature.

\begin{acknowledgments}
The authors thank Drs. I.G. Siny and O.E. Kvyatkovskii for helpful discussions. The work 
was partially supported by RFBR Grant 02-02-17678, and President RF ss-1416.2003.2, 
and NSF-DMR-FY2004 grants. 
\end{acknowledgments}

%

\end{document}